\documentclass[runningheads]{llncs}
\usepackage[T1]{fontenc}
\usepackage{graphicx}
\usepackage{amsmath}
\usepackage{comment}
\usepackage{tikz}
\usepackage{tikz-cd}
\usepackage{float}
\usepackage{placeins}
\usetikzlibrary{shapes.geometric, arrows, positioning, arrows.meta, calc, backgrounds, matrix, shapes.symbols, shapes.misc, decorations.markings, shadows, decorations.pathreplacing, backgrounds}
\pgfdeclarelayer{background}
\pgfsetlayers{background,main}
\makeatletter
\renewenvironment{abstract}{%
      \list{}{\advance\topsep by0.00cm\relax\small
      \leftmargin=0.0cm 
      \rightmargin=\leftmargin}\item[\hskip\labelsep
                                     \bfseries\abstractname]}
     {\endlist}
\makeatother

\begin{document}
\title{The Environmental Cost of LLMs in AIED: Reporting and Practices \thanks{Paper Accepted for Blue Sky Track of 27th International Conference on AI in Education (2026)}}

%
%

\author{Sabrina C. Eimler \inst{1}\orcidID{0000-0001-8944-2814} \and
Lukas Erle\inst{1}\orcidID{0000-0001-8623-8869} \and 
Daniel Flood\inst{2}\orcidID{0000-0002-2834-3473} \and
Aditi Haiman\inst{3}\orcidID{1111-2222-3333-4444} \and
Luca Häckert\inst{1}\orcidID{2222--3333-4444-5555} \and
André Helgert \inst{1}\orcidID{0000-0001-6008-4793} \and
Lachlan McGinness\inst{4}\orcidID{0000-0002-3231-4827}\and
Büsra Yapici\inst{1}\orcidID{2222--3333-4444-5555}
}
\authorrunning{S. Eimler et al.}
%
\institute{Institute of Computer Science and Institute of Positive Computing, Ruhr West University of Applied Sciences,
Lützowstraße 5, 46236 Bottrop, Germany 
\and
Centre for Computational Science and Mathematical Modelling, Coventry University, Coventry, United Kindgom, CV1 5FB\\
 \and
Carnegie Mellon University\\
\and 
Australian National University and CSIRO
 }
%
\maketitle              
\begin{abstract}

Large Language Model (LLM) usage in recent years has become increasingly widespread in the Artificial Intelligence in Education (AIED) community. While LLMs offer unique 
avenues for learners and educators, using LLMs comes with computational and environmental costs. 
These costs are
mostly hidden due to a lack of standardised 
procedures to measure and report these impacts. To address this gap, we first conducted a literature review of all $N = 396$ papers published as part of the AIED 2025 conference proceedings, determining if and how computational or environmental costs of LLMs are reported. 
Most 
projects use LLMs, 
but few report computational resources used and almost none discuss environmental impacts of LLMs as an ethical concern.  

To address this lack of standardised reporting practices, we propose an open-source method for systematically measuring and reporting the computational expense of LLMs and environmental impact of running Machine Learning (ML) AIED systems. 
We provide software solutions to measure the carbon footprint for both local and cloud based hardware. We also provide an easy-to-use formula to calculate the computational expense of frontier LLMs even when the exact number of parameters is not known.  Overall, we hope to motivate colleagues to use our method to strive for more transparent reporting of hidden costs of using LLMs in the AIED community.


\end{abstract}

\subsection*{Sustainability Metrics for this Paper}
\small{
Total Corrected Carbon Intensity: $1.55\times10^{-3}$kg \\
Total Computational Cost:  $2.60\times10^{10}$ FLOPs\\
Minimum Training Carbon required to reach 100\% (IRIS UCI Dataset): $1.823\times10^{-6}$kg \\
Training Efficiency ($k$) for Breast Cancer UCI Dataset:  $2.04\times 10^{6}$ kg$^{-1}$ of CO$_2$\\
}
\normalsize

\section{Introduction}

In recent years, Large Language Models (LLMs) have become widely used in many aspects of academic research. In the education space, we have seen generative AI become embedded in tutoring agents, writing assistants, assessment tools, automated grading and feedback systems, and teacher supports, among many others \cite{rismanchian2025evolution}. 
Within the Artificial Intelligence in Education (AIED) community, LLM-based systems are increasingly used as learning and research tools. This rapid integration of generative AI tools represents an important shift in the AIED community, as educational Artificial Intelligence (AI) increasingly involves LLMs that are computationally expensive.  As LLMs become normalised in AIED research, it is important to consider questions about methodology and ethical issues.

Much of the conversation about LLMs in AIED has focused primarily on their effectiveness as learning tools, implications of biases and accuracy, student learning outcomes, data privacy, and responsible use \cite{holmes2022aied}. Some of the main ethical concerns include fairness and transparency (as related to issues of training data and plagiarism) and student over-reliance on LLMs. While these issues are incredibly important, the computational and environmental costs of using LLMs are largely ignored in the AIED community. 

Training and deploying LLMs requires substantial computational resources that have environmental impacts \cite{Henderson2020Energy,Luccioni2024Power,morrison2025environmentalimpact} that are still underestimated. Even when researchers access models through APIs or through the web rather than training them directly, the infrastructure required for inference is still energy-intensive. As the goal of educational tools is to benefit students at large, the scale of LLM usage in education means increasingly large computational demands. However, perhaps because LLMs are still relatively new, there is no standardised practice to report computational usage. Environmental considerations are rarely discussed as an important ethical consideration in the AIED community. 

Our work focuses on two questions:
\begin{enumerate}
    \item How does the AIED community currently report the computational and environmental costs of LLM usage in their research?
    \item How can we encourage transparency and accuracy in estimating environmental costs in the research of the AIED community?
\end{enumerate}

To address these questions, we first reviewed all of the papers from the 2025 AIED conference proceedings. We looked at how LLMs are used (if at all), whether or not computational specifics are reported, and whether environmental sustainability is discussed as an ethical concern. We use the findings of this review to provide a current picture of the usage of LLMs in the AIED community and highlight the lack of attention given to computational infrastructure in educational AI research. 
Next, we propose a several methods that researchers can use to calculate the carbon emissions created by LLMs and Machine Learning (ML) methods in research projects. 
We describe an existing open-source tool (CodeCarbon) for estimating carbon emissions, create an open-source wrapper to generalise to a variety of sustainability metrics, and present a method for theoretically estimating the computational resources used by LLMs. 
We argue that awareness of environmental issues and how LLMs contribute should become standardised in AIED through norms for reporting specifics of LLM usage in papers. By making the environmental footprint of LLMs more visible, the community can set standards for sustainable educational AI research.

\section{LLMs in the AIED Community}

To understand the frequency and type of use of LLMs in the AIED community, we carefully reviewed all 396 papers from the AIED 2025 conference proceedings. This included all types of submissions, including full and short papers, workshops and tutorials, the doctoral consortium, the BlueSky track, practitioners, industry and policy (PIP), and late breaking results. We have explicitly limited ourselves to the proceedings of the most recent conference, as this gives us a clearly defined and most up-to-date snapshot of the work carried out within the AIED community that reflects the current handling of LLMs and their (often implicit) computational and environmental costs. Additionally, because LLM-based systems have only recently been widely integrated into AIED research, it is likely that tools and practices for transparency and reporting of computational and environmental costs are only just emerging.

\vspace{2mm}

In order to systematically record how LLMs are treated in AIED research, we searched the full text of each paper for specific information regarding the use of LLMs. 
For each paper, we recorded the following:

\begin{enumerate}

\item \textbf{LLM usage (yes/no)}: A binary indicator as to whether an LLM was used or discussed as part of the system, method, or evaluation (beyond conceptual mentions).
\item \textbf{Type of LLM usage (categorisation):} We have classified LLM usage into four roles. The categories were identified using a mix of deductive and inductive approaches:
 \begin{enumerate}
    \item Subject of Experiment - Evaluating the capability of the LLM.
    \item Data Analysis - The LLM is used in analysing quantitative or qualitative data (eg. labeling images or coding qualitatitve data).
    \item Design of Products - The LLM is used to generate content or design part of a product that will be presented to a user (eg. generating test questions for teachers).
    \item User Interaction - The LLM is used to interact directly with a research participant such that the LLM is the main product of the project (eg. a conversational tutor or AI assistant).
 \end{enumerate}
\item \textbf{Description of LLM usage (description):} A short  description of what the LLM is used for.
\item \textbf{Reporting of computational costs (yes/no):} Whether the paper provides any details on computational costs (e.g., runtime, inference cost, energy usage, hardware type, or similar).
\item \textbf{LLM usage details (model/version):} Whether the paper specifies which models were used, including any further details (e.g., GPT-4o, Claude, Llama).
\item \textbf{Mention of sustainability or environmental issues (yes/no):} Whether the papers mentions environmental impact, sustainability, energy use or similar topics.

\end{enumerate}

If a paper used multiple LLMs or if a model fits into multiple roles, we coded all applicable roles individually. This coding framework allows us to quantify how frequently LLMs appear in AIED 2025 and how they are used. To increase methodological rigor, the coding scheme was iteratively developed and applied by two coders, with disagreements discussed and resolved by consensus. Ambiguous and borderline cases were documented during the coding process and used to refine category definitions to improve consistency across papers. 

\subsection{Results of the Literature Review}

Out of the 396 identified AIED 2025 papers, 257 have used an LLM and 139 have not.
There was a variety of LLM usage across papers. 
100 papers use LLMs as subjects in experiments, mainly for systematic comparison of models and prompting strategies, and evaluating bias, robustness and tutoring performance. 
79 papers use LLMs for data analysis and automation. The tasks included coding, annotation, and evaluation of dialogues, essays and code logs. Some studies also used LLMs for video/audio transcription, extraction of structured features, and summarisation.
108 papers used LLMs for content design, mainly creating tasks, hints, distractions, lesson plans, and learning materials. 
The most common use of LLMs was as a component in an interactive learning system. 176 papers reported using LLMs as tutors to provide hints and feedback to students, to simulate a training system or as a coding and writing assistant.
These four roles often occur in combination and are sometimes not clearly separated.

Only 85 papers provided information on computational cost, and sustainability issues were only mentioned in 57 papers. 
In contrast, the LLM usage specifications are often specifically named. Among the models used, OpenAI models (especially GPT-4/4o) appear most frequently, as well as Anthropic Claude and Google Gemini. 
In addition, open source LLMs such as LLaMA, Mistral, Gemma, Qwen and DeepSeek are also used, some with information on parameter sizes and fine-tuning. In addition, related NLP/speech/multi-modal components are also mentioned (Whisper, BERT).

This review shows that 
while some work has at least mentioned environmental impact, the topic is not mainstream.
Almost every paper that does mention the computational cost or environmental impact of LLM usage reports these considerations in a different way. 
Some papers differ in which metrics are reported, some offer strategies for reducing the impact, and some just peripherally mention sustainability as a possible drawback of widespread adoption of system. 
This paints a picture of a community which lacks standardised tools and procedures to report and evaluate the environmental impact of LLM usage. 
In the following sections, we outline specific procedures to measure and systematically report the sustainability of different forms of AI tools.

\section{Measuring Computational Cost of Local Hardware}
\label{sec:CodeCarbon}

As shown in the literature review, the AIED community lacks standardised reporting for the environmental cost of AI systems. 
In this section, we highlight CodeCarbon, an open-source tool that can be used to estimate the carbon emissions of running AI tools. 

The environmental impact of an algorithm can be calculated by measuring direct energy usage from hardware and converting it to CO$_2$ emissions based on the energy mix of the local electricity grid \cite{lottick2019energy}. 
Tools such as CodeCarbon calculate this by querying hardware interfaces 
to measure instantaneous wattage over the duration of the computational process \cite{lottick2019energy,codecarbon_docs}.
CodeCarbon is an open-source, lightweight Python package that measures the power draw of local hardware and uses assumptions about the local electricity grid to estimate the carbon intensity (amount of CO$_2$ emitted) over a specified duration.

When low-level hardware measurements are unavailable, CodeCarbon assumes that the hardware is using 50\% of its Thermal Design Power (TDP), the maximum amount of heat that the cooling system CPU or GPU is designed to dissipate \cite{codecarbon_docs}.  
CodeCarbon does this by detecting the hardware used by the local machine and looking up the TDP in internally stored tables. If local hardware is unlisted, global constants (e.g., 85W for a CPU) are applied instead \cite{codecarbon_docs}.

We propose that CodeCarbon or equivalent systems could be used to calculate the total power consumed and CO$_2$ released by research when it is published for an AIED conference or journal. 
This could be as simple as stating ``Total $\text{CO}_2$ estimate: $1.2\times10^{-3}$kg'' after the abstract but before the introduction to make the reader aware of the carbon intensity of the research.

Kocher et al. demonstrated that CodeCarbon's dynamic tracking of hardware usually underestimates total energy consumption by approximately 20\% \cite{Kocher2025EnergyNAS}. This deficit occurs because software tracking cannot profile power supply unit inefficiencies, hardware cooling systems, and peripheral devices \cite{lottick2019energy}. 

Although CodeCarbon and equivalent tools are not perfectly accurate, they can be used to provide a rough estimate of the carbon intensity of any local AIED system. 
We encourage members of the AIED community to report the carbon intensity values of their research not only to raise awareness but also to make this a quantifiable metric that the community actively works to improve.
The exact metric that should be reported depends strongly on the AI technique that is being used. 
For example, for an AI tutoring system, it may be most appropriate to report the carbon intensity per minute of tutoring for one student. For an automated grading system, it would be more appropriate to report the carbon intensity per question (or per page) graded.

In the following section, we present methods that can be used to improve the accuracy of the reported values by introducing the concept PUE-equivalent for local machines. 
We also present other more nuanced measures that could be used to measure the sustainability of ML techniques.
Furthermore, we introduce a proof-of-concept tool that evaluates the performance of standard ML models in light of the computational and environmental costs of their training.

\section{A Proposed Framework for Evaluating the Sustainability of Machine Learning}
\label{sec:ML_Evaluation}

In the previous section, we showed how the open-source tool CodeCarbon could be used to determine the CO$_2$ emissions from running a local AI system. 
This default carbon intensity metric does not account for differences in hardware utilisation\footnote{Hardware utilisation is the fraction of a system's total processing capacity (such as CPU cycles, GPU cores, or memory bandwidth) that is engaged in executing a workload at a given moment.} between local devices and cloud servers. 
Servers maintain high utilisation rates across their hardware. 
In contrast, local development environments operate with lower average hardware utilisation. 
A local device draws baseline power for the operating system and display regardless of the ML workload. 

To quantify these overheads, data centers use the Power Usage Effectiveness (PUE) metric, defined as the ratio of total facility energy to the energy consumed strictly by the computing hardware, see Equation \ref{eq:PUE}.

\begin{equation}
\label{eq:PUE}
\text{PUE} = \frac{\text{Total Facility Energy}}{\text{IT Equipment Energy}}
\end{equation}

We suggest that the concept of PUE values could be extended to calculate a ratio of the total energy consumed by a local computer over the energy used by its computational hardware.
This accounts for energy that is not used directly for computation (for example, to cool CPUs and to power monitors), but could be considered wasted when calculating the total computational cost of an AI or ML task.
We will refer to this generalised ratio for local hardware as PUE-equivalent.

Because CodeCarbon 
only measures the energy used by the computational subsystem of the machine, it effectively defaults to a PUE-equivalent of 1.0. 
As a result, local energy measurements may not accurately represent the footprint of models when deployed in local production environments.

Besides this limitation of CodeCarbon, another challenge faced by the AIED community is the lack of a standard way to evaluate the trade-off between predictive performance and energy consumption. 
Machine learning research traditionally reports metrics such as accuracy, but rarely contextualises these results in terms of computational cost. 
Without a shared framework, it is difficult to compare models that achieve similar performance but differ substantially in energy usage.


To address these issues, we propose a set of metrics that could be used to evaluate the efficiency of local hardware and report the accuracy of models in light of their computational cost. 
We have developed a proof-of-concept wrapper for underlying tracking tools (such as CodeCarbon).
We have named our system \texttt{TerraFlops} \footnote{See \url{https://github.com/danflood94/TerraFlops}}, note the capitalisation is different from TerraFLOPs (Terra FLoating Point Operations) and TerraFLOPS (Terra FLoating Point Operations per Second) to avoid confusion. 
\texttt{TerraFlops} integrates raw hardware emissions data with facility-level PUE variables and ML performance metrics to generate a unified sustainability score.
This tool was built to explore methods for combining environmental and performance metrics, and we present it here to share this approach with the AIED community. 

Firstly, \texttt{TerraFlops} enables a researcher to estimate the cost of deployment on Cloud hardware through its \texttt{cloud} mode. This should be used when researchers develop models locally but intend to calculate deployment costs at scale. \texttt{TerraFlops} contains PUE values for different providers which can be accessed via flags such as \texttt{AWS}, \texttt{GCP}, or \texttt{GENERIC}. This overrides the default 1.0 PUE by applying fixed PUE variables based on documented cloud provider averages.
This allows researchers developing on consumer-grade hardware to output a standardised projection of their model's data centre cost, independent of local utilisation inefficiencies.


For local diagnostics, the framework includes a \texttt{local\_auto} mode which calculates a dynamic PUE based on real-time hardware load. Workloads utilising less than 10\% of the system capacity are assigned a PUE of 1.60 to account for the baseline power overhead of the local device, while high-utilisation workloads scale down to a PUE of 1.08. These values may still be an under-estimate, as research suggests the energy efficiency (T-FLOPs per joule) of consumer-grade GPUs is significantly lower than more efficient cloud-based hardware \cite{McGinness2025Can}. Accurately refining these benchmarks remains an area for future work, and the wrapper is designed to be easily adjusted by users to reflect their specific hardware.

\subsection{Composite Performance and Sustainability Metrics}
In this section, we present metrics that utilise the PUE-equivalent values to report environmental impact. 
Typically, the most prominently reported metrics for ML models are accuracy-based metrics, and as a result, ML models are typically optimised for predictive power. 
In order to encourage the community to optimise model efficiency and minimise environmental impact, we propose the prominent reporting of sustainability metrics at the start of a paper. 
The total carbon intensity calculated by CodeCarbon can be reported for any paper that evaluates an AIED system.
Additionally, we propose three composite metrics to quantify the environmental cost of performance gains of ML Models:

\begin{enumerate}
    \item \textbf{Corrected Carbon Intensity:} The framework retrieves the raw carbon intensity calculated by CodeCarbon \cite{codecarbon_zenodo} and multiplies this value by the PUE-equivalent determined by the current flag. 
    \item \textbf{Carbon per Accuracy:} When training an ML model, the carbon intensity of the training is divided by the model's performance metric (e.g., Accuracy or Macro F1 Score). 
    \item \textbf{Sustainability Score:} Carbon intensity values vary by many orders of magnitude between lightweight algorithms and deep neural networks. If we would like to provide a simple `Sustainability Score' between 1 and 10, a linear scale will be insufficient for cross-model comparison. Therefore, we demonstrate how non-linear functions (base-10 logarithmic) could be applied to the \texttt{Carbon\_per\_Accuracy} ratio to create a user-friendly 10-point scale.
\end{enumerate}

\texttt{Corrected Carbon Intensity} takes into consideration hardware utilisation. 
Training a model over a long period of time at a low hardware load results in a higher proportion of wasted baseline overhead. 
By applying the dynamic PUE-equivalent penalty, the framework captures this inefficiency and penalises models run on underutilised local hardware.

\texttt{Carbon per Accuracy} can be interpreted as the amount of CO$_2$ emissions required to achieve an accuracy of 1.0 (100\%) under the (false) assumption that accuracy scales linearly with computational expense. 
In reality, the performance of ML models plateaus during training, and the relation between accuracy and training compute is highly non-linear.
This means that carbon per accuracy will likely reward under-trained models and not reward strong performance.

One way to improve this metric in future implementation 
is to apply 
non-linear functions to the accuracy, to appropriately reward improvements. 
This non-linear accuracy could be divided by the carbon intensity to give a more nuanced carbon efficiency metric.
For example, for training tasks which approach 100\% accuracy in the limit of infinite training, it could be assumed that accuracy ($A$) takes the following functional form: 
\begin{equation}
A = 1 - e^{(-kt)}    
\end{equation}
Where $k$ is a training constant which indicates the efficiency/sustainability of training and $t$ is a measure of computational expense (either FLOPs, Joules or CO$_2$ emissions).
Simple algebraic rearrangement gives:
\begin{equation}
k = \frac{\log_e(1-A)}{t}    
\end{equation}
This training efficiency value could be reported as a fairer measure than carbon per accuracy.
We believe that an important area for future work is refining these assumptions to determine the fairest methods for determining sustainability of training and testing.


Our implementation includes an example of how a 1-10 \texttt{Sustainability Score} could be calculated, but additional work will be required to determine a universal standard score that is general enough to be applied to any ML application. 

For illustrative purposes, we utilised the \texttt{TerraFlops} wrapper when training a logistic regression (ML) model on the Iris and Breast Cancer datasets from the UCI machine learning repository \cite{UCIdataset}. We provide the sustainability statistics immediately before the introduction as an example of how these metrics could be reported. 

This proof-of-concept tool will allow members of the AIED community to report efficiency and sustainability metrics in addition to accuracy. 
It will also allow the community to take into account the computational cost of new ML methods and evaluate whether small gains in accuracy are worthwhile in light of the environmental impact.
In the upcoming section, we shift our focus from traditional ML classification to LLMs. 
We provide an easy way for members of the AIED community to calculate and report the cost of running local LLMs and provide a method that could be used to roughly estimate the computational cost of running proprietary models.

\section{Estimating Computational Expense of LLMs}
\label{sec:LLMFlops}

Measuring the carbon intensity of hardware can be an accurate method of estimating environmental impact. 
However, it may not be the most equitable system to compare the computational cost across countries. 
Carbon intensity unfairly advantage researchers with access to more efficient hardware, or who live in countries with a `cleaner' power grid.
A more equitable measure of AI usage is the computational cost.

\begin{figure}[ht]
	\centering
\newcommand{\dimdmodel}{2.0cm} 
\newcommand{\dimdattn}{1.5cm} 
\newcommand{\dimdff}{5cm}
\newcommand{\dimnvocab}{11cm}
\newcommand{\dimdone}{0.8cm}
\scalebox{0.75}{
\begin{tikzpicture}[
		node distance=1.5cm and 2cm,
		boxstyle/.style={
			draw, 
			thick, 
			fill=blue!5, 
			text width=3.5cm, 
			minimum height=1.2cm, 
			align=center
		},
		arrow/.style={-stealth, thick},
		boxEmbeddingMatrix/.style={
			draw, thick, fill=green!5, 
			text width=\dimnvocab, 
			minimum height=\dimdmodel, 
			align=center
		},
		boxRMSNormLayer/.style={
			draw, thick, fill=red!5, 
			text width=\dimdmodel+2cm, 
			minimum height=\dimdone, 
			align=center
		},
		boxAttentionWeightMatrix/.style={
			draw, thick, fill=blue!5, 
			text width=\dimdattn, 
			minimum height=\dimdmodel, 
			align=center
		},
		boxAttentionOutputMatrix/.style={
			draw, thick, fill=blue!5, 
			text width=\dimdmodel, 
			minimum height=\dimdmodel, 
			align=center
		},
		boxFeedForwardMatrix/.style={
			draw, thick, fill=blue!5, 
			text width=\dimdff, 
			minimum height=\dimdmodel, 
			align=center
		},
		clean/.style={
			draw=none,      
			fill=none,      
			text width=5cm, 
			align=center,   
			inner sep=2pt   
		},
		flowArrow/.style={
			-stealth, 
			blue, 
			text=blue,
			font=\small,
			shorten <= 5pt, 
			shorten >= 5pt
		},
		]

		\node[boxEmbeddingMatrix] (n1) {Token Embedding Matrix};
		\draw[<->, thin, gray] 
		([yshift=5pt]n1.north west) -- ([yshift=5pt]n1.north east) 
		node[midway, above, font=\small] {$n_{\text{vocab}}$};
		\draw[<->, thin, gray] 
		([xshift=-5pt]n1.north west) -- ([xshift=-5pt]n1.south west) 
		node[midway, above, font=\small, rotate=90] {$d_{\text{model}}$};
		
		\node[boxRMSNormLayer] (n2) [below=1.0of n1] {RMS Norm Layer ($\gamma$)};
		\draw[<->, thin, gray] 
		([yshift=5pt]n2.north west) -- ([yshift=5pt]n2.north east) 
		node[midway, above, font=\small] {$d_{\text{model}}$};
		\draw[<->, thin, gray] 
		([xshift=-5pt]n2.north west) -- ([xshift=-5pt]n2.south west) 
		node[midway, left, rotate=0] {$1$};

		\coordinate (rowCenter) at ([yshift=-5.0cm]n2.south);
		\node[blue, align=center, font=\small] at (rowCenter) {Scale and \\ Softmax: \\ $5.5n_{\text{heads}}n_{\text{ctx}}$\\ $\times(n_{\text{ctx}}+1)$};
		
		\node[boxAttentionWeightMatrix] (n32) [left=1.1of rowCenter] {Query Weights ($W_{Q_j}$)};
		\begin{pgfonlayer}{background}
		\node[boxAttentionWeightMatrix] (n32backback) at ([xshift=-4mm, yshift=4mm]n32.center) {Query Weights $W_{Q_j}$};
		\node[boxAttentionWeightMatrix] (n32back) at ([xshift=-2mm, yshift=2mm]n32.center) {Query Weights $W_{Q_j}$};
		\end{pgfonlayer}
		\draw[<->, thin, gray] 
		([yshift=5pt]n32backback.north west) -- ([yshift=5pt]n32backback.north east) 
		node[midway, above, font=\small] {$d_{\text{model}}$};
		\draw[<->, thin, gray] 
		([xshift=-5pt]n32backback.north west) -- ([xshift=-5pt]n32backback.south west) 
		node[midway, above, font=\small, rotate=90] {$d_{\text{model}}$};

		\node[boxAttentionWeightMatrix] (n31) [left=2.5of n32] {Key Weights ($W_{K_j}$)};
		\begin{pgfonlayer}{background}
			\node[boxAttentionWeightMatrix] (n31backback) at ([xshift=-4mm, yshift=4mm]n31.center) {Key Weights $W_{K_j}$};
			\node[boxAttentionWeightMatrix] (n31back) at ([xshift=-2mm, yshift=2mm]n31.center) {Key Weights $W_{K_j}$};
		\end{pgfonlayer}
		\draw[<->, thin, gray] 
		([yshift=5pt]n31backback.north west) -- ([yshift=5pt]n31backback.north east) 
		node[midway, above, font=\small] {$d_{\text{model}}$};
		\draw[<->, thin, gray] 
		([xshift=-5pt]n31backback.north west) -- ([xshift=-5pt]n31backback.south west) 
		node[midway, above, font=\small, rotate=90] {$d_{\text{model}}$};

		\node[boxAttentionWeightMatrix] (n33) [right=1.9of rowCenter] {Query Weights ($W_{V_j}$)};
		\begin{pgfonlayer}{background}
		\node[boxAttentionWeightMatrix] (n33backback) at ([xshift=-4mm, yshift=4mm]n33.center) {Value Weights $W_{V_j}$};
		\node[boxAttentionWeightMatrix] (n32back) at ([xshift=-2mm, yshift=2mm]n33.center) {Value Weights $W_{V_j}$};
		\end{pgfonlayer}
		\draw[<->, thin, gray] 
		([yshift=5pt]n33backback.north west) -- ([yshift=5pt]n33backback.north east) 
		node[midway, above, font=\small] {$d_{\text{model}}$};
		\draw[<->, thin, gray] 
		([xshift=-5pt]n33backback.north west) -- ([xshift=-5pt]n33backback.south west) 
		node[midway, above, font=\small, rotate=90] {$d_{\text{model}}$};

		\node[boxAttentionOutputMatrix, yshift=4mm] (n34) [right=1.0of n33] {Output Weights ($W_{O_j}$)};
		\draw[<->, thin, gray] 
		([yshift=5pt]n34.north west) -- ([yshift=5pt]n34.north east) 
		node[midway, above, font=\small] {$d_{\text{model}}$};
		\draw[<->, thin, gray] 
		([xshift=-5pt]n34.north west) -- ([xshift=-5pt]n34.south west) 
		node[midway, above, font=\small, rotate=90] {$d_{\text{model}}$};

		\draw[decorate, decoration={brace, amplitude=10pt}, thick]
		($(n31.north west) + (-0.6, 1.7)$) -- ($(n34.north east) + (0, 1.3)$)
		node[midway, above=8pt] {Multi-head attention};

		\node[boxRMSNormLayer] (n4) [below=3.0of rowCenter] {RMS Norm Layer ($\gamma$)};
		\draw[<->, thin, gray] 
		([yshift=5pt]n4.north west) -- ([yshift=5pt]n4.north east) 
		node[midway, above, font=\small] {$d_{\text{model}}$};
		\draw[<->, thin, gray] 
		([xshift=-5pt]n4.north west) -- ([xshift=-5pt]n4.south west) 
		node[midway, left, rotate=0] {$1$};
		
		\coordinate (rowCenterff) at ([yshift=-2.5cm]n4.south);

		\node[boxFeedForwardMatrix] (n51) [left=0.5of rowCenterff] {Feed Forward Network \\ Gate Projection ($W_{\text{gate}}$)};
		\draw[<->, thin, gray] 
		([yshift=5pt]n51.north west) -- ([yshift=5pt]n51.north east) 
		node[midway, above, font=\small] {$d_{\text{ff}}$};
		\draw[<->, thin, gray] 
		([xshift=-5pt]n51.north west) -- ([xshift=-5pt]n51.south west) 
		node[midway, above, font=\small, rotate=90] {$d_{\text{model}}$};

		\node[boxFeedForwardMatrix] (n52) [right=0.5of rowCenterff] {Feed Forward Network \\ Up Projection ($W_{\text{up}}$)};
		\draw[<->, thin, gray] 
		([yshift=5pt]n52.north west) -- ([yshift=5pt]n52.north east) 
		node[midway, above, font=\small] {$d_{\text{ff}}$};
		\draw[<->, thin, gray] 
		([xshift=-5pt]n52.north west) -- ([xshift=-5pt]n52.south west) 
		node[midway, above, font=\small, rotate=90] {$d_{\text{model}}$};

		\node[clean] (n6) [below=of rowCenterff]{};
		\node[blue, align=center, font=\small, yshift=-20pt] (activation) [left=1.0 of n6] {Activation Function: \\ $n_A n_{\text{ctx}} d_{\text{ff}}$};

		\node[boxFeedForwardMatrix] (n7) [below=1.7of n6] {Feed Forward Network \\ Down Projection ($W_{\text{down}}$)};
		\draw[<->, thin, gray] 
		([yshift=5pt]n7.north west) -- ([yshift=5pt]n7.north east) 
		node[midway, above, font=\small] {$d_{\text{ff}}$};
		\draw[<->, thin, gray] 
		([xshift=-5pt]n7.north west) -- ([xshift=-5pt]n7.south west) 
		node[midway, above, font=\small, rotate=90] {$d_{\text{model}}$};
		
		\draw[decorate, decoration={brace, amplitude=10pt, mirror}, thick]
		($(n2.north west) + (-7.5, 0.5)$) -- ($(n7.south west) + (-7.0, 0)$)
		node[midway, left=15pt, rotate=90, anchor=south] {Attention Layer/ Transformer Layer ($\times n_{\text{layers}})$};

		\node[boxRMSNormLayer] (n8) [below=1.0of n7] {RMS Norm Layer ($\gamma$)};
		\draw[<->, thin, gray] 
		([yshift=5pt]n8.north west) -- ([yshift=5pt]n8.north east) 
		node[midway, above, font=\small] {$d_{\text{model}}$};
		\draw[<->, thin, gray] 
		([xshift=-5pt]n8.north west) -- ([xshift=-5pt]n8.south west) 
		node[midway, left, rotate=0] {$1$};
		
		\node[boxEmbeddingMatrix] (n9) [below=0.8of n8] {Token Embedding Matrix \\
		(Same as first matrix for architectures which use tied input-output embeddings.)};
		\draw[<->, thin, gray] 
		([yshift=5pt]n9.north west) -- ([yshift=5pt]n9.north east) 
		node[midway, above, font=\small] {$n_{\text{vocab}}$};
		\draw[<->, thin, gray] 
		([xshift=-5pt]n9.north west) -- ([xshift=-5pt]n9.south west) 
		node[midway, above, font=\small, rotate=90] {$d_{\text{model}}$};
		

		
		\draw[flowArrow, shorten <= 5pt, shorten >= 5pt] 
		($(n1.south west)!0.75!(n1.south east)$)
		|-
		node[midway, right, yshift=0.5cm, text width=3cm] {$\text{Layer normalisation}: 4n_{\text{ctx}}d_{\text{model}}$}
		(n2.east);
		
		\draw[flowArrow, shorten <= 5pt, shorten >= 10pt] 
		($(n2.south west)!0.1!(n2.south east)$) --
		++(0,-7pt) -|
		node[midway, above, yshift=0.0cm, text width=3cm] {$\text{Calculation of Key matrices}: 2n_{\text{ctx}}d_{\text{model}^2}/g$}
		($(n31backback.north) + (-0pt, 10pt)$);
		
		\draw[flowArrow, shorten <= 5pt, shorten >= 10pt] 
		($(n2.south west)!0.25!(n2.south east)$) -- 
		++(0,-12pt) -| 
		node[midway, left, yshift=-1.0cm, text width=2.5cm, align=center] {Calculation of \\ Query matrices: \\ $2n_{\text{ctx}}d_{\text{model}}^2$}
		($(n32backback.north) + (-5pt, 10pt)$);
		
		\draw[flowArrow, shorten <= 5pt, shorten >= 10pt] 
		($(n2.south west)!0.75!(n2.south east)$) -- 
		++(0,-12pt) -| 
		node[midway, right, yshift=-1.0cm, text width=2.5cm, align=center] {Calculation of \\ Query matrices: \\ $2n_{\text{ctx}}d_{\text{model}}^2/g$}
		($(n33backback.north) + (5pt, 10pt)$);
		
		\draw[flowArrow, shorten <= 10pt, shorten >= 15pt] 
		($(n31backback.north west)!0.1!(n31.south west)$) --
		++(-30pt,-0pt) |-
		node[midway, below, yshift=0.0cm, text width=3cm, align=center] {RoPE: \\ $3n_{\text{ctx}}d_{\text{model}}/g$}
		($(n31backback.north west)!0.7!(n31.south west)$);
		
		\draw[flowArrow, shorten <= 10pt, shorten >= 15pt] 
		($(n32backback.north west)!0.1!(n32.south west)$) --
		++(-30pt,-0pt) |-
		node[midway, below, yshift=0.0cm, text width=3cm, align=center] {RoPE: \\ $3n_{\text{ctx}}d_{\text{model}}$}
		($(n32backback.north west)!0.7!(n32.south west)$);
		
		\draw[flowArrow, shorten <= 5pt, shorten >= 15pt] 
		($(n31.south west)!0.5!(n31.south east)$) --
		++(-0pt,-20pt) -|
		node[midway, below, xshift=-4.0cm, yshift=0.0cm, text width=5cm, align=center] {$Q_jK_j^\top$:  $n_{\text{ctx}}(n_{\text{ctx}}+1)d_{\text{model}}$}
		($(rowCenter) + (0pt, -10pt)$);
		
		\draw[flowArrow, shorten <= 5pt, shorten >= 15pt] 
		($(n32.south west)!0.5!(n32.south east)$) --
		++(-0pt,-20pt) -|
		node[midway, below, xshift=-3.0cm, yshift=0.0cm, text width=3cm, align=center] {}
		($(rowCenter) + (0pt, -10pt)$);
		
		\draw[flowArrow, shorten <= 25pt, shorten >= 15pt] 
		($(rowCenter) + (0pt, 5pt)$) 
		|-
		node[midway, above, xshift=-0.0cm, yshift=0.0cm, text width=3cm, align=center] {Multiplication by $V_j$ \\ $d_{\text{model}} n_{\text{ctx}}$\\ $\times (n_{\text{ctx}}+1)$}
		($(n33backback.north west)!0.1!(n33backback.south west)$);
		
		\draw[flowArrow, shorten <= 10pt, shorten >= 15pt] 
		($(n33.south west)!0.5!(n33.south east)$) -- 
		++(0,-20pt) -- 
		++(30pt,0) |- 
		node[midway, below, xshift=-2.5cm, yshift=-1.4cm, text width=5cm, align=right] {Multiplication by $W_{O_j}$ \\ $2n_{\text{ctx}}d_{\text{model}}^2$}
		($(n34.north west)!0.9!(n34.south west)$);
		
		\draw[flowArrow, shorten <= 5pt, shorten >= 5pt] 
		($(n1.north east)!0.5!(n1.south east)$) --
		++(50pt,0pt)  |-
		node[midway, left, xshift=+0.7cm, yshift=5.0cm, text width=3cm, align=center] {Add \\ embedding: \\ $n_{\text{ctx}}d_{\text{model}}$}
		($(n34.north east)!0.8!(n34.south east)$);
		
		\draw[flowArrow, shorten <= 5pt, shorten >= 5pt] 
		($(n34.south west)!0.8!(n34.south east)$)
		|-
		node[midway, left, xshift=0.0cm, yshift=2.0cm, text width=3cm, align=right] {Layer \\ normalisation: \\ $4n_{\text{ctx}}d_{\text{model}}$}
		($(n4.north east)!0.5!(n4.south east)$);
		
		\draw[flowArrow, shorten <=2pt, shorten >= 10pt] 
		($(n4.south west)!0.25!(n4.south east)$) --
		++(0,-7pt) -|
		node[midway, left,  xshift=1.2cm, yshift=-0.4cm, text width=5cm, align=left] {Multiplication by $W_\text{gate}$: $2n_{\text{ctx}}d_{\text{model}}d_{\text{ff}}$}
		($(n51.north) + (-0pt, 10pt)$);
		
		\draw[flowArrow, shorten <=2pt, shorten >= 10pt] 
		($(n4.south west)!0.75!(n4.south east)$) --
		++(0,-7pt) -|
		node[midway, right,  xshift=0.4cm, yshift=-0.4cm, text width=5cm, align=left] {Multiplication by $W_\text{up}$: $2n_{\text{ctx}}d_{\text{model}}d_{\text{ff}}$}
		($(n52.north) + (-0pt, 10pt)$);
		
		\draw[flowArrow, shorten <=2pt, shorten >= 10pt] 
		($(n51.south west)!0.1!(n51.south east)$) --
		node[midway, right,  xshift=0.4cm, yshift=-0.4cm, text width=5cm, align=left] {}
		($(activation) + (-1pt, 10pt)$);
		
		\draw[flowArrow, shorten <=10pt, shorten >= 10pt] 
		($(activation) + (40pt, 0pt)$) -|
		node[midway, above,  xshift=-2.2cm, yshift=-0.0cm, text width=5cm, align=left] {Elementwise Multiplication:\\ $n_{\text{ctx}}d_{\text{ff}}$}
		($(n7.north west)!0.85!(n7.north east)$);
		
		\draw[flowArrow, shorten <=2pt, shorten >= 10pt] 
		($(n52.south west)!0.255!(n52.south east)$) --
		node[midway, right,  xshift=0.0cm, yshift=-0.0cm, text width=5cm, align=left] {Multiplication by $W_\text{down}$: $2n_{\text{ctx}}d_{\text{model}}d_{\text{ff}}$}
		($(n7.north west)!0.85!(n7.north east)$);
		
		\draw[flowArrow, shorten <= 5pt, shorten >= 4pt] 
		($(n7.south west)!1.00!(n7.south east)$)
		|-
		node[midway, right, yshift=0.5cm, text width=5cm, align=left] {Layer normalisation\\ Final token only:\\ $4d_{\text{model}}$}
		(n8.east);
		
		\draw[flowArrow, shorten <= 3pt, shorten >= 7pt] 
		($(n8.south west)!0.05!(n8.south east)$)--
		++(0,-7pt)	-|
		node[midway, left, xshift=3.0cm, yshift=0.7cm, text width=7cm, align=left] {Multiplication of final token by embedding \\ matrix and application of final softmax to \\ result to get probability distribution \\ $2d_{\text{model}}n_{\text{vocab}} + 10n_{\text{vocab}}$}
		($(n9.north west)!0.05!(n9.north east)$);
\end{tikzpicture}
}
	\caption{Diagram of decoder-only transformer architecture demonstrating the number of parameters and FLOPs required for the first token generated by a Large Language Model after receiving input context. Grey arrows indicate the matrix dimensions that can be used to find the number of learned parameters within the model. Blue arrows indicate the number of FLOPs required to generate the first token.}
	\label{fig:transformer}
\end{figure}
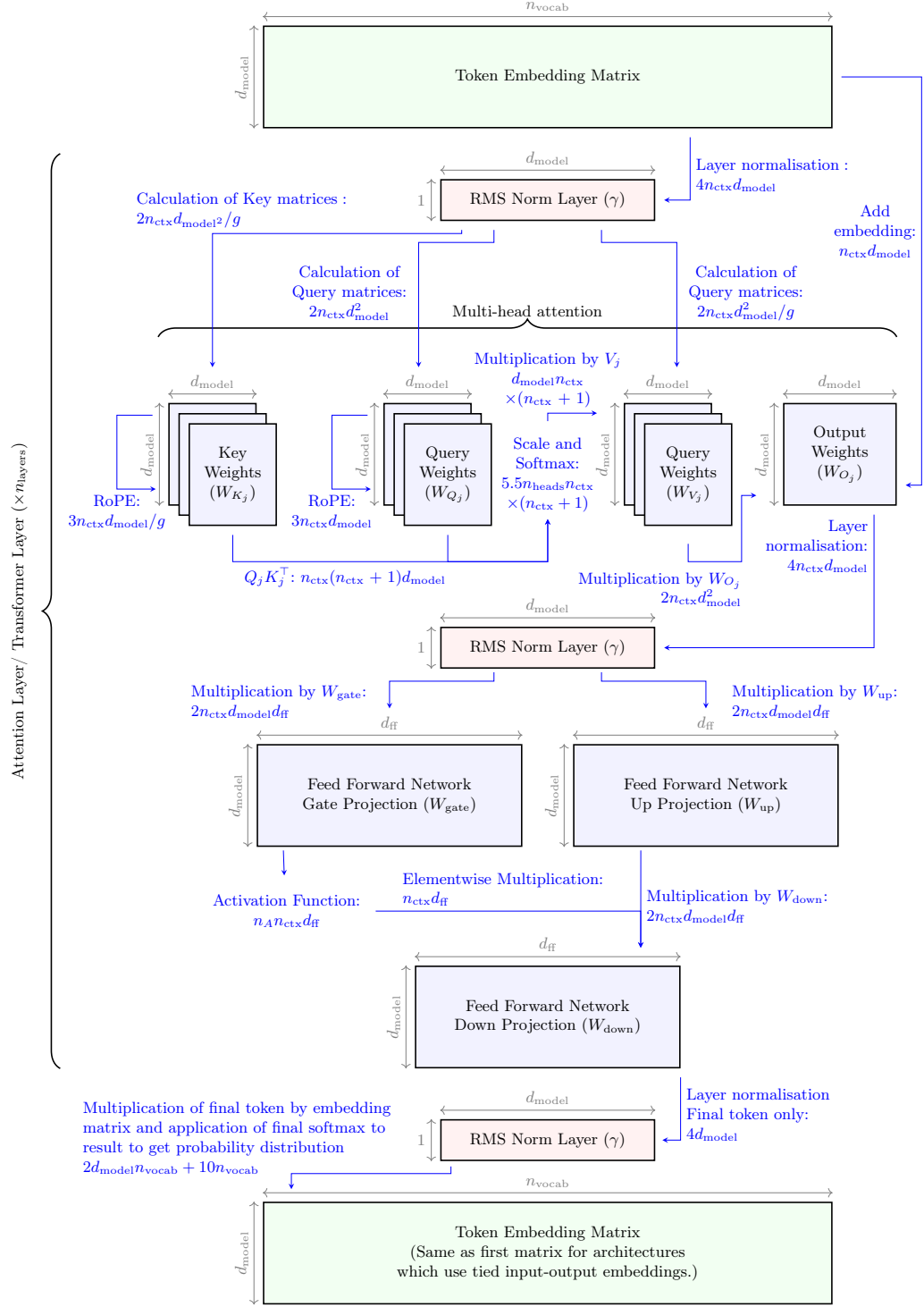

In this section, we outline the theoretical computational cost of the most common architecture of open source LLMs: decoder-only transformers. 
If we know the architecture of the transformer and the number of prompt (input) tokens ($n_{\text{ctx}}$) and completion (output) tokens ($n_{\text{output}}$), then we can calculate the theoretical number of floating-point operations (FLOPs) required to complete the task. 
Figure \ref{fig:transformer} visually represents the transformer.

Grey arrows indicate the matrix dimensions that can be used to calculate the number of active parameters in each component. Blue arrows indicate the number of FLOPs required to generate the first token. 
As long as sufficient memory is available, subsequent tokens are computationally cheaper because of the KV-caching mechanism which stores the Key and Value matrices from previous calculations \cite{Dai2019Transformer}.

In determining the theoretical FLOPs, we assume the Pre-Layer Normalisation decoder-only transformer. \footnote{We use the following notation for the model parameters:
$d_{\text{model}}$ is the model dimension/embedding size,  $d_{\text{ff}}$ is the expanded dimension of the feed forward layer, $d_{\text{attn}}$ is the dimension of each attention head, given by $d_{\text{model}} = n_{\text{heads}} d_{\text{attn}}$, $n_{\text{heads}}$ is number of attention heads in each multi-head attention block, $n_{\text{layer}}$ is the number of layers, $n_{\text{A}}$ is the number of FLOPs per activation function, and $n_{\text{vocab}}$ is the number of tokens in the vocabulary. We assume that the implementation requires 10 FLOPs per entry in a softmax function.}

Additionally, we assume that the model considered uses Rotary Position Embeddings (RoPE) \cite{Su2024RoPE}, RMS Layer Normalisation \cite{Zhang2019RMSNorm}, and KV caching, which are common for most open source LLMs released in 2025.

We note that recent versions of DeepSeek reduce the cost and number of parameters through Multi-Headed Latent Attention, but we do not consider this in our calculations \cite{DeepSeek2024DeepSeekV2}.
We also assume that the computational cost of multiplying two matrices of dimensions $a\times b$ and $b \times c$ is $2abc$ FLOPs \cite{Kaplan2020Scaling}. The only exception to this is the attention mechanism, where we assume that the upper diagonal of the matrix does not need to be calculated and therefore requires $ba(a+1)$ FLOPs. Therefore, as illustrated in Figure \ref{fig:transformer}, the total cost of the first token is:
\scriptsize
\begin{flalign*}
	C_1 = &n_{\text{layer}}[8n_{\text{ctx}}d_{\text{model}}^2+
	11n_{\text{ctx}}d_{\text{model}}+ 
	2n_{\text{ctx}}(n_{\text{ctx}}+1)d_{\text{model}} +
	5.5 n_{\text{heads}} n_{\text{ctx}}(n_{\text{ctx}}+1)] + \\ 
    &n_{\text{layer}} (5n_{\text{ctx}} d_{\text{model}} + 
	6 n_{\text{ctx}} d_{\text{model}} d_{\text{ff}}+
	(n_A+1)n_{\text{ctx}} d_{\text{ff}}) +
	4d_{\text{model}} + 2d_{\text{model}}n_{\text{vocab}} + 10n_{\text{vocab}}\\
\end{flalign*}
\normalsize

\FloatBarrier

\noindent The cost of subsequent tokens ($i >1$) is:
\scriptsize
\begin{flalign*}
	C_i = & 
	n_{\text{layer}}[11d_{\text{model}}+8d_{\text{model}}^2+4d_{\text{model}}(n_{\text{ctx}}+i) + 11n_{\text{heads}}(n_{\text{ctx}}+i)] + \\ 
	&n_{\text{layer}} (5d_{\text{model}} + 6d_{\text{model}} d_{\text{ff}}+(n_A+1)d_{\text{ff}}) +
	4d_{\text{model}}+ 2d_{\text{model}}n_{\text{vocab}} + 10n_{\text{vocab}}\\
\end{flalign*}
\normalsize

We anticipate that for most users, the difficulty of finding each of the required parameters to use the formula could be prohibitive. 

\FloatBarrier

There is a simpler approximation for the total FLOPs for an LLM call, which in previous studies has been found to be accurate within 10\%  \cite{Kaplan2020Scaling,McGinness2025Can}, see Equation \ref{eqn:FLOP_approximation}.

\begin{equation}
\label{eqn:FLOP_approximation}
    \text{FLOPs} = 2Nn
\end{equation}

Where $N$ is the number of active parameters in the model, and $n$ is the total number of prompt and completion tokens. 
Using this formula will make it significantly easier for authors to estimate the computational cost of running open source models. 

However, we are aware that many members of the AIED community will use proprietary models run on external servers. 
In this case, the authors will be unable to use CodeCarbon to measure the carbon intensity, and the researchers will not know the number of active parameters of the LLMs that they are using.
In this situation we recommend the following method for reporting computational cost and environmental impact:
\begin{enumerate}
    \item Sum the number of context and output tokens ($n$). Most APIs for LLM inference charge users based on the number of input and output tokens and therefore readily report these numbers to the user.
    \item Estimate the number of active parameters ($N$) of the LLM being used. For proprietary LLMs, the exact number of parameters is not publicly available. However, if we assume that frontier models are a similar size to open source models with similar performance we can make a rough, order of magnitude estimate. 
    Based on the architecture of Qwen3, state-of-the-art models likely have approximately 100 billion active parameters, while flash/fast models most likely have 30 billion active parameters \cite{Yang2025Qwen3}. 
    We note that according to Equation \ref{eqn:FLOP_approximation}, the approximated FLOPs are linear with respect to model size. Therefore if this model size estimate is underestimated by a multiplicative factor, so the reported FLOPs will be underestimated by the same factor. 
    \item Use Equation \ref{eqn:FLOP_approximation} to estimate the number of FLOPs for the AIED system being tested. Then, if we assume that inference is served on an H100 GPU with a Global mean PUE of 1.5, the energy cost would be approximately $2$ T-FLOPs/Joule \cite{McGinness2025Can}. If we assume that the servers are based in the United States where approximately $384$ grams of CO$_2$ are released per KWh \cite{Ember2026CarbonIntensity}, this gives an approximate value of $6\times10^{-5}$ grams of CO$_2$ per T-FLOP.
\end{enumerate}
Although this three-step procedure only gives an order of magnitude estimate of FLOPs and carbon intensity, we would encourage researchers to use this approximation as a consistent sustainability benchmark that can be used for any LLM-based AI system.

\section{Conclusions and Recommendations}

The integration of Large Language Models (LLMs) and other Artificial Intelligence (AI) systems into educational technology requires accountability regarding their environmental impact. 
A review of all 396 papers from AIED 2025 found that the majority (257) of papers explicitly use LLMs, while only a fraction (57) consider their environmental impact.
Even among the few papers that mention sustainability, there are no standardised reporting techniques used, making comparison between papers difficult. 

We present three recommendations for standardised reporting sustainability of AIED research. 
Firstly, for any AIED system run on local or cloud hardware we recommend using CodeCarbon to measure and report carbon intensity (details in Section \ref{sec:CodeCarbon}).
Secondly, for Machine Learning tasks, we propose a framework for the reporting of carbon emissions alongside accuracy using our \texttt{TerraFlops} wrapper. 
This implements a method to calculate PUE-equivalence for local hardware and reports three composite metrics that will focus the AIED community on performance relative to CO$_2$ output (Section \ref{sec:ML_Evaluation}).
Thirdly, we recommend using theoretical FLOP estimations to quantify the computational cost and CO$_2$ released from proprietary LLMs (Section \ref{sec:LLMFlops}). 

To further improve the framework, the PUE-equivalent baselines could be refined to consider the performance of consumer-grade GPUs compared to cloud-based GPUs. 
Another limitation of our current framework is that it does not consider the water consumption of cloud-based GPUs, which has a very significant detrimental effect on some communities. 
Additionally, specific non-linear relations could be developed for the \texttt{Carbon per Accuracy Metric} to correctly account for performance plateaus during classical ML training.
Finally, both CodeCarbon and \texttt{TerraFlops} should be tested for non-expert usability, and a consensus should be developed for a 1-10 sustainability score for AIED researchers who work in different sub-disciplines.

In order to increase buy-in from researchers, future AIED conferences and competitions could require a reporting of sustainability metrics as a condition of submission, similar to anonymity requirements. 
By adopting these practices, the AIED community can systematically evaluate, monitor, and reduce the environmental cost of AI in education, while maintaining transparency about the sustainability of AI research.


 \subsubsection{Acknowledgements} 
This work was partly supported by the Transferhub project, which is funded by the Ministry of Economic Affairs, Industry, Climate Action and Energy of the State of North Rhine-Westphalia and the European Union as part of the NRW 2021-2027 EFRE/JTF program [EFRE20800649].

\subsubsection{Author Contributions}
B\"usra Yapici, Luca H\"ackert, Andr\'e Helgert, and Lukas Erle conducted the literature review, with the latter two authors analyzing the compiled literature and writing the chapter. Sabrina C. Eimler supervised the literature review. 
Aditi Haiman contributed to Literature review methodology, and writing/editing of the manuscript. Daniel Flood created the \texttt{TerraFlops} Repository and wrote chapters 3 and 4 and contributed to editing. Lachlan McGinness performed the FLOPs analysis, refined the data analysis methodology of \texttt{TerraFlops} and contributed editing. 

%
%
%
%

\end{document}